\documentclass[final,1p,times]{elsarticle}
\usepackage{lineno,hyperref}
\usepackage{color}
\usepackage{graphicx}
\usepackage{amssymb}
\usepackage{float}
\usepackage{amsmath}
\biboptions{sort&compress}
\modulolinenumbers[5]
\usepackage{mwe}
\usepackage{subfig}

\journal{Optics Communications}

\begin{document}

\begin{frontmatter}

\title{Spectral singularity enhances transverse spin}

\author[a]{Mounica Mahankali}

\author[a,b]{Sudipta Saha}
\author[a]{Nirmalya Ghosh}

\author[c,d]{S. Dutta Gupta \corref{mycorrespondingauthor}}
\cortext[mycorrespondingauthor]{Corresponding author}
\ead{sdghyderabad@gmail.com}
\address[a]{Department of Physical Sciences, Indian Institute of Science Education and Research Kolkata, Mohanpur-741 246, India}
\address[b]{Department of Biomedical Engineering, Florida International University, Miami, FL 33174, USA}
\address[c]{Tata Centre for Interdisciplinary Sciences, TIFRH, Hyderabad 500107, India}
\address[d]{School of Physics, University of Hyderabad, Hyderabad-500046, India}

\begin{abstract}
We study the transverse spin and the Belinfante spin momentum in a gain-loss balanced waveguide, known to be one of the first and most studied examples of $\mathcal{PT}$-symmetric systems. Such a guide supports the spectral singularities leading to infinite scattering amplitudes for both reflection and transmission. We show that near the spectral singularity there can be dramatic enhancement of the transverse spin and the transverse spin momentum. Note that these exotic spin and spin momentum have recently been observed experimentally despite having tiny magnitudes, making it worthy to explore ways and means to enhance these fundamentally important elusive quantities. 
\end{abstract}

\begin{keyword}
Transverse spin, Belinfante spin momentum, $\mathcal{PT}$-symmetric waveguide.
\end{keyword}

\end{frontmatter}


\section{Introduction}
In recent years there has been a great deal of interest in the angular momentum carried by light \cite{allenbook,sdgbook,bliokhphyrep}. It is now well understood that in addition to usual longitudinal angular momentum structured light can possess an extraordinary transverse spin angular momentum (SAM) \cite{bliokh2014,aiello2015,bekshaev2015,bauer2016,zayats2014,measuring2015,aiello2016,encyclopedia2018}. While the transverse SAM is helicity independent, the transverse momentum can be sensitive to the helicity, opening up new possibilities for spin optics \cite{bliokh2014,aiello2015,bekshaev2015}. The existence of such nonstandard spin-momentum was first discussed by Belinfante in a seminal field theoretical paper in 1940 \cite{belinfante1940}. The spin momentum discovered by Belinfante was considered to be virtual or elusive, since it has no effect on dipolar particles (in the sense of transferring energy or exerting optical pressure). In contrast, the usual canonical momentum exerts the radiation pressure leading to easily observable experimental effects \cite{berry2009,bliokh2014}. Despite the tiny effect produced by this extraordinary spin momentum, it has now been observed experimentally \cite{banzer2013,measuring2015,natphys_2016}. There have been studies pointing to the novel use in spin locking and mapping fields in nano-optical systems \cite{bliokh2015,zubinoptica,zubinapl,nanoscale2018,garoli2017helicity}. The generic feature responsible for the extraordinary spin and spin-momentum is the existence of a longitudinal component of the field in structured light \cite{bliokhphyrep,bliokh2014}. Note that the conventional featureless plane waves do not have any such components and thus cannot exhibit the unusual effects. In contrast, strongly focused beams \cite{banzer2013} or evanescent waves associated with total internal  reflection or surface excitations of plasmonic systems can have very strong longitudinal components of field \cite{bliokh2014,bliokh2012}. The other essential aspect is the finite phase difference between longitudinal and transverse components of the field vector resulting in a `polarization ellipse' in the plane of incidence \cite{bliokh2015}. The rotating tip of the field vector in time in the plane of incidence results on transverse spin. All these features have been discussed at great length in several recent reviews \cite{bliokhphyrep,encyclopedia2018,a3,bekshaev2018}. One of the major directions that emerge is how to enhance the tiny Belinfante momentum. Strong coupling mediated dispersion management was proposed to be one of the possible ways \cite{saha2018} to enhance the transverse spin. Coherent perfect absorption \cite{wan2011,deb2007} was shown to further enhance the effects \cite{samyo}. Only passive systems were probed. To the best of our knowledge, there are no  studies on transverse spin and spin momentum in optical systems with gain, keeping this in view in this paper we propose a gain-loss balanced $\mathcal{PT}$-symmetric waveguide for enhancing the transverse spin and spin-momentum. The proposed system serves several goals. Firstly, the wave-guiding geometry ensures the existence of a longitudinal field component with the necessary phase lag as compared to the transverse component. Second and most importantly, a properly tuned $\mathcal{PT}$-symmetric guide can have a spectral singularity, leading to blown up scattering amplitudes for both reflection and transmission. Though infinities are not encountered in realistic nonlinear systems \cite{sdgpt, sdgptol}, there is always a significant field enhancement. We show that the resonances of `null' width of the $\mathcal{PT}$-symmetric system can lead to dramatic enhancement of the transverse spin. Recall that the interest in the $\mathcal{PT}$-symmetric systems was initially motivated by mathematical and fundamental aspects that concerned real spectra of non-Hermitian systems \cite{bender1998}. Since then a great deal of research has been carried out on photonic systems \cite{zhao2018,philos2013}. There are now several studies highlighting few practical applications in photonics. A very recent study exploits the highly dispersive properties of such systems for stopping and storing light \cite{goldzak2018}. However, $\mathcal{PT}$-symmetry has not been exploited so far in the context of the elusive Belinfante's spin-momentum.
\begin{figure}[ht]
	\centering
	\includegraphics[width=0.7\linewidth]{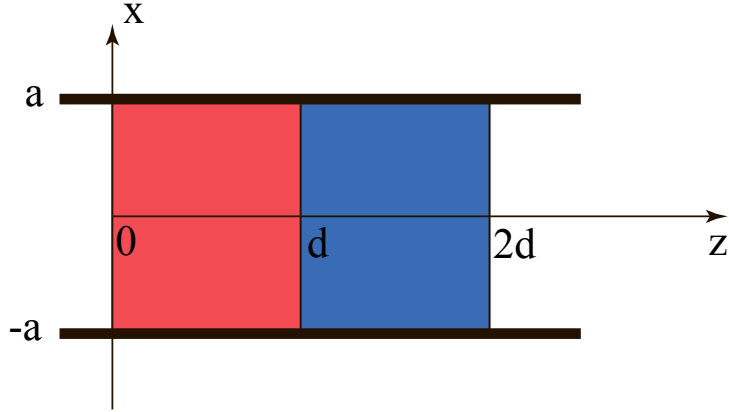}
	\caption{Schematic of the waveguide geometry}  
	\label{fig1}
\end{figure}
\par
\section{Formulation of the problem}

Consider the balanced waveguide of width $2a$ having equal sections of gain and loss media each of length $d$ (see Fig. \ref{fig1}). The guide has an infinite extent along $y$ so that we can ignore the variations along $y$ ($\frac{\partial}{\partial y}=0$). Outside the gain-loss region the guide is filled with a dielectric with dielectric constant $\bar\epsilon_i=1$, while the dielectric function of the gain /loss medium is given by
\begin{equation}
\label{eq1}
\bar\epsilon_{1,2}=1+\zeta_{1,2}\frac{\omega_p^2}{(\omega^2-\omega_0)^2+2i\gamma\omega},
\end{equation}
where $\omega_p$ and $\omega_0$ are the plasma and the resonance frequencies, respectively,  and $\gamma$ gives the decay constant.   The subscript 1 (2) refer to the gain (loss) medium with $\zeta_1=+1$ ($\zeta_2=-1$). Throughout the text, permittivity and relative permittivity are denoted as \(\epsilon\) and \(\bar\epsilon\) while permeability and relative permeability are denoted as \(\mu\) and \(\bar\mu\), respectively. 
\par
Let the guide be excited from the left by a TM mode with the only non-vanishing magnetic field component along $y$. As mentioned in the Introduction, such a guide satisfies the necessary conditions of $\mathcal{PT}$-symmetry, since Eq. (\ref{eq1}) fulfills the requirement for the system-wide variation of the dielectric function $\epsilon(z-d)=\epsilon^*(d-z)$ near $\omega=\omega_0$. In what follows we examine the electric and magnetic field in each region and outline the procedure to arrive at the spin and spin momentum densities.
\subsection{Fields, spin and spin momentum density}
We follow electric-magnetic democracy and use scaled fields $\tilde{\mathbf{E}}=\sqrt{\epsilon_0}\mathbf{E}$ and $\tilde{\mathbf{H}}=\sqrt{\mu_0}\mathbf{H}$ since they have the same dimension \cite{berry2009,sdgbook}. The presence of a waveguide restricts the \(x\)-component of the wavevector as $k_x = \frac{\pi}{2a}$. The only non-vanishing component of the magnetic field in different regions can be expressed as follows
\begin{align}
\tilde{H}_{0}&=\left(A_{i}e^{ik_{0z}z}+A_{r}e^{-ik_{0z}z}\right)\cos\left(k_xx\right),\hspace{60pt} \rm{for}~ z<0, \label{eq2}\\
\tilde{H}_{1}&=\left(A_{1+}e^{ik_{1z}z}+A_{1-}e^{-ik_{1z}z}\right)\cos\left(k_xx\right),\hspace{45pt} \rm{for}~ 0<z<d, \label{eq3}\\
\tilde{H}_{2}&=\left(A_{2+}e^{ik_{2z}(z-d)}+A_{2-}e^{-ik_{2z}(z-d)}\right)\cos\left(k_xx\right),\hspace{10pt} \rm{for}~d<z<2d, \label{eq4}\\
\tilde{H}_{t}&=A_{t}e^{ik_{0z}(z-2d)}\cos\left(k_xx\right),\hspace{110pt} \rm{for}~ z>2d, \label{eq5}
\end{align}
In Eqs.(\ref{eq2}) -- (\ref{eq5}), subscripts 0 and t refer to the medium of incidence and transmittance, while 1 (2) points to the gain (loss) medium occupying $0<z<d$ ($d<z<2d$). $k_x$ and $k_{jz}=\sqrt{k_0^2\bar\epsilon_j-k_x^2}$, $(j=0,1,2,t)$ are the  $x$ and $z$ components of the wave vector, respectively. Using Eqs. (\ref{eq2}) -- (\ref{eq5}) one can easily calculate the corresponding electric fields. For example,
\begin{align}
\tilde E_{jx} &= p_{jz}Q_{j-}\cos\left(k_xx\right),\label{eq6}\\ 
\tilde E_{jz} &= \frac{-ip_x}{\bar\epsilon_j}Q_{j+}\sin\left(k_xx\right),\label{eq7}
\end{align}
where we have introduced the following abbreviated notations
\begin{align}
Q_{j+}&=A_{j+}e^{ik_{jz}(z-z_j)}+A_{j-}e^{-ik_{jz}(z-z_j)}, \label{eq8}\\
Q_{j-}&=A_{j+}e^{ik_{jz}(z-z_j)}-A_{j-}e^{-ik_{jz}(z-z_j)}.\label{eq9}
\end{align}
Here,  $p_{jz} = k_{jz}/k_0\bar\epsilon_j$, ($j = 1,2$) and $p_x = k_x/k_0$. Demanding the continuity of the tangential fields $\tilde{E}_x ~\textrm{and} ~\tilde{H}_y$ across the interfaces (i.e., $z = 0,~d,~2d$), one can calculate the amplitudes in all the media. For example,
\begin{align}
\begin{pmatrix}
A_i\\A_r\end{pmatrix}&=\begin{pmatrix}1&1\\p_{0z}&-p_{0z}\end{pmatrix}^{-1}M_1~ M_2\begin{pmatrix}1\\p_{0z}
\end{pmatrix}A_t,\label{eq10}
\end{align}
where \(M_1~\textrm{and}~M_2\) are the characteristic matrices of gain and loss media, respectively \cite{sdgbook}.
\begin{figure}[ht]
	\centering	\includegraphics[width=0.8\linewidth]{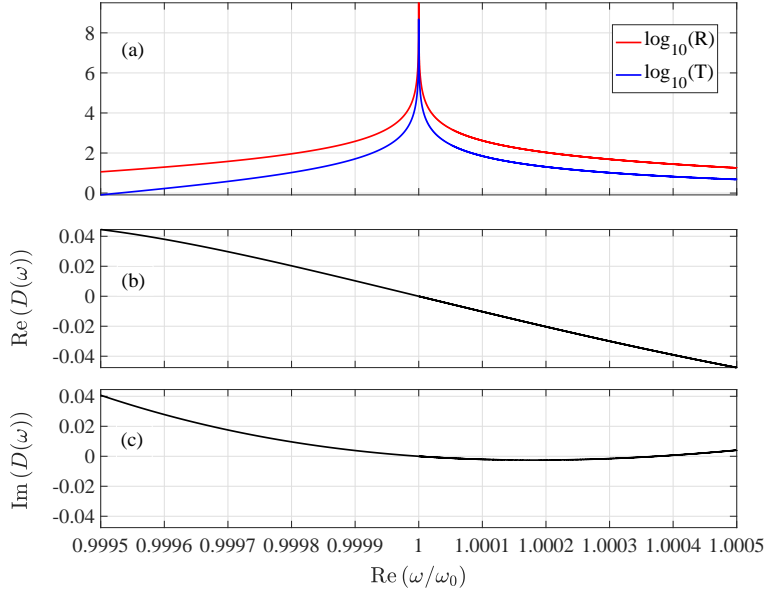}
	\caption{(a) R and T as a function of Re($\omega/\omega_0$); (b) Re(D($\omega$)) and (c)Im(D($\omega$)) as a function of Re($\omega/\omega_0$) for $a_0 = 0.062045958076$ $\mu$m, $d_0=1.004358293913$ $\mu$m, $\omega_0=5 eV$, the decay constant $\gamma=1.25 eV$ and plasma frequency $\omega_p=0.2 eV$ }
	\label{fig2}
\end{figure}
\par
As mentioned in the Introduction, a great deal of research has been carried out on the structure of the Poynting current and optical momentum density. it is now clear that the momentum density in a monochromatic optical field can be decomposed into two parts, namely, the orbital (canonical) momentum $\mathbf{p_o}$ and `intrinsic' spin momentum $\mathbf{p_s}$ as follows \cite{bliokhphyrep,berry2009,bliokh2014,aiello2015,a3,a5,berry2015} 
\begin{align}
\mathbf{p}&=\mathbf{p_o}+\mathbf{p_s}, \label{eq11}\\
\mathbf{p_s}&=\frac{1}{2}(\nabla\times\mathbf{s}),\label{eq12}\\ 
\mathbf{s}&=\frac{1}{4\omega}\textrm{Im}(\bar\epsilon\mathbf{\tilde E^*}\times\mathbf{\tilde E}+\bar\mu\mathbf{\tilde{H}^*}\times\mathbf{\tilde{H}})\label{eq13},
\end{align}
where \(\mathbf{s}\) is the spin density. Since, \(\mathbf{\tilde H}\) has a single component, it will not contribute to the spin density. Simplifying Eq. (\ref{eq13}), we get
\begin{equation}
\mathbf{s} = \frac{1}{2\omega}\textrm{Re}(\bar\epsilon)\textrm{Im}(\tilde E_x\tilde E_z^*)\hat{y}, \label{eq14}
\end{equation}
Substituting Eqs. (\ref{eq6}) and (\ref{eq7}) in Eqs. (\ref{eq12}) and (\ref{eq13}), we get
\begin{align}
\mathbf{s}&=\frac{\textrm{Re}(\bar\epsilon_j)p_x}{4\omega}\sin\left(2k_xx\right)\textrm{Re}\left(\frac{p_{jz}}{\bar\epsilon_j^*}Q_{j+}^*Q_{j-}\right), \label{eq15}\\
\mathbf{(p_s)}_z&=\frac{\textrm{Re}(\bar\epsilon_j)p_xk_x}{4\omega}\cos(2k_xx)\textrm{Re}\left(\frac{p_{jz}}{\bar\epsilon_j^*}Q^*_{j+}Q_{j-}\right),\label{eq16}\\
\mathbf{(p_s)}_x&=\frac{\textrm{Re}(\bar\epsilon_j)p_x}{8\omega}\sin(2k_xx)\left|Q_{j+}\right|^2\textrm{Im}\left(\frac{p_{jz}k_{jz}}{\bar\epsilon^*_j}\right) \label{eq17}.
\end{align}
\subsection{Spectral singularity and its effect on the spin density and spin momentum density}
From Eq.(\ref{eq10}), it is straight forward to define the amplitude (intensity) reflection $r$ ($R$) and transmission $t$ ($T$) coefficients, respectively, as follows
\begin{align}
\label{eq18} \
r = \frac{A_r}{A_i},\hspace{30pt}t = \frac{A_t}{A_i},\hspace{30pt}\mathbf{R} = \left|r\right|^2,\hspace{30pt}\mathbf{T}=\left|t\right|^2. 
\end{align}
\par
A close inspection of Eq. \ref{eq18} reveals that both $r(\omega)$ and $t(\omega)$ have a common denominator $D(\omega)$ and the dispersion relation for the modes is given by \cite{sdgbook}
\begin{equation}
D(\omega)= (m_{11}+m_{12}p_f)p_i+(m_{21}+m_{22}p_f)=0,
\label{eq19}
\end{equation}
where $m_{ij}$ are the elements of the total characteristics matrix.
The dispersion relation has a complex root in general. The real part of the root of this equation $Re(\omega)$ gives the location of the resonance, while the imaginary part $Im(\omega)$ embodies the losses of the mode. A spectral singularity corresponds to a resonance of vanishing width rendering the root of dispersion relation on the real axis with $Im(\omega)=0$. This in turn implies that $D(\omega)$
as a function of real $\omega$ must be zero at the spectral singularity. In the next section we reproduce some of the well known results on spectral singularity in the context of our system and carry forward the calculations for extracting the spin momentum and spin densities. We highlight the existence of the elusive and unusual transverse (perpendicular to the direction of the waveguide axis) components of both these quantities. We show that the spectral singularity can lead to dramatic enhancements of both these extraordinary features. 
\section{Numerical results and discussions}

In what follows we present the numerical results pertaining to the balanced gain/loss guide. We start with the well known feature of the spectral singularity for such guides \cite{muga,mostafazadeh}. The spectral singularity for both $R$ and $T$ along with the real and imaginary parts of $D(\omega)$ are shown in Fig. \ref{fig2} (a), (b) and (c), respectively. The singularities occur at $\omega=\omega_0$ and for very specific values of system parameters. Note that the response given by Eq.\ref{eq1} is PT-symmetric only at $\omega=\omega_0$. As discussed in \cite{mostafazadeh} the singularities can be labeled by an integer. We show one of such resonances in Fig \ref{fig2}a for $\omega/\omega_0=1$ and for $a=a_0=0.062045958076$, $d=d_0=1.004358293913$. Henceforth we will label the parameters corresponding to the singularity by subscript 0. As can be seen from Fig. \ref{fig2}a, both $R$ and $T$ blow up as a consequence of the vanishing width resonance. The plot of $D(\omega)$ as a function of real $\omega$ confirms that. Indeed, the vanishing of both real and imaginary parts of $D(\omega)$ for real $\omega$ at the spectral singularity confines the root of the dispersion relation Eq.\ref{eq19} on the real axis. It is clear that such a root points to a resonance of infinitely narrow width. 
\par
\begin{figure}[ht]
	\centering
	\includegraphics[width=0.9\linewidth]{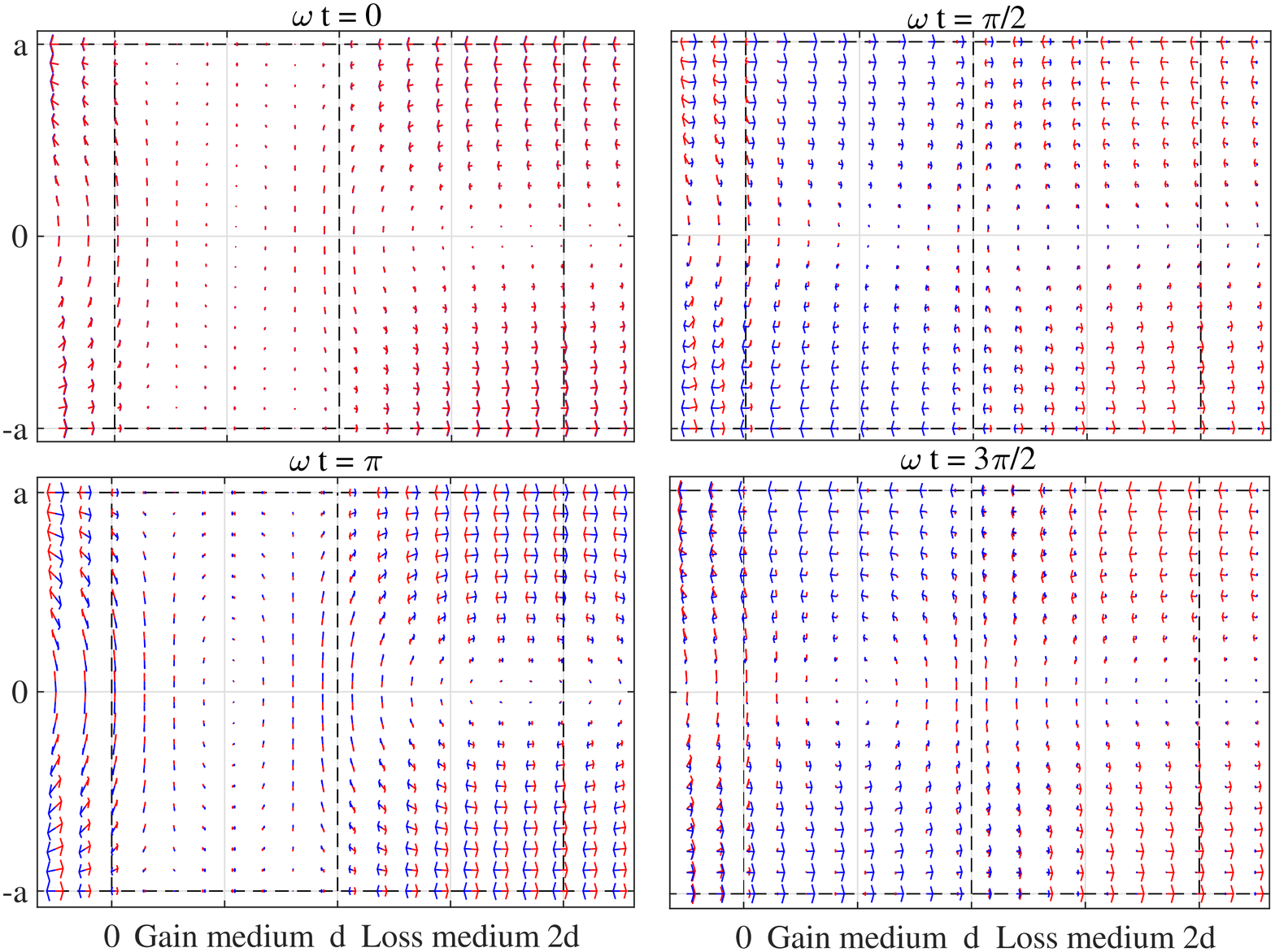}
	\caption{Quiver plot of Electric field throughout the waveguide at $\omega t = 0$ (red) and $\omega t = 0, \pi /2, \pi, 3\pi /2$ where $\omega = \omega_0= 0.062045958076 \mu$m, $a = a_0=0.062045958076$ $\mu$m and $d = d_0=1.004358293913$ $\mu$m, the decay constant $\gamma=1.25 eV$ and plasma frequency $\omega_p=0.2 eV$}
	\label{fig3}
\end{figure}
It is clear from Eqs.\ref{eq14} and \ref{eq15} that the spin density has only one non-vanishing component along y axis due to the TM character of the waveguide mode. We use Eq.\ref{eq15} to calculate the spin density for both the resonant and off-resonant cases. Recall that for the typical parameters used in our system one of the spectral singularities occurs for $a_0=0.062045958076$ $\mu$m, $d_0=1.004358293913$ $\mu$m and $\omega/\omega_0=1$. Rotation of the field vector has been identified as the source of transverse spin and spin momentum \cite{saha2018,bliokh2014}. Analogous circulation is shown
in Fig.\ref{fig3}, where the snap shots of the electric field vector is plotted at different times, namely, at $\omega t=0,\pi/2,\pi,3\pi/2$. A comparison of these panels clearly reveal the rotation of the electric field vector in the plane of incidence ($x-z$ plane). In fact, this circulation leads to the transverse spin along the $y$ direction in full conformity with the earlier results \cite{bliokh2014,saha2018}. 
\begin{figure}[ht]
		\begin{minipage}{.5\linewidth}
		\centering
		\includegraphics[width=\linewidth]{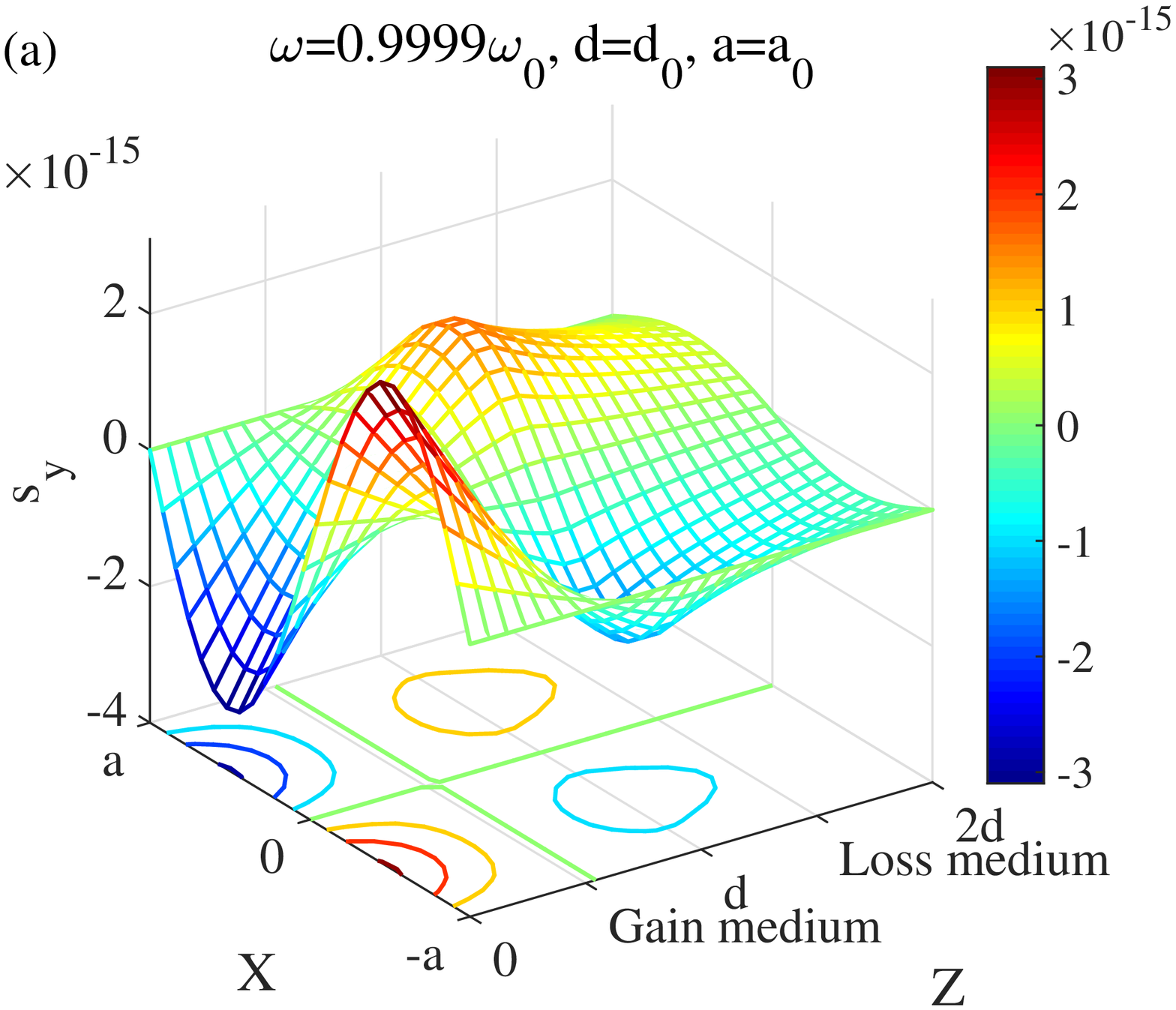}
	\end{minipage}%
	\begin{minipage}{.5\linewidth}
		\centering
			\includegraphics[width=\linewidth]{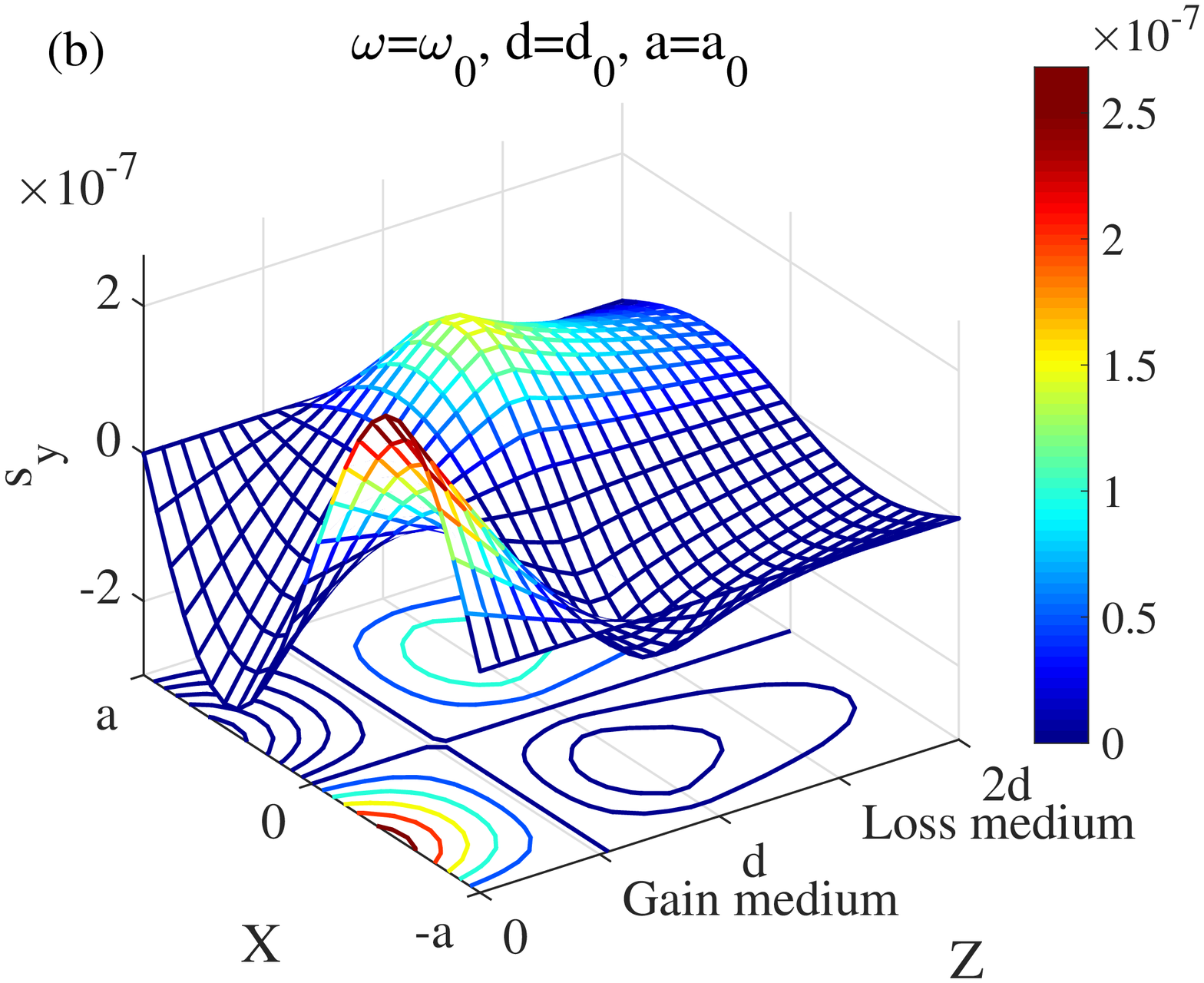}
	\end{minipage}\par\medskip
	\centering
		\includegraphics[width=0.5\linewidth]{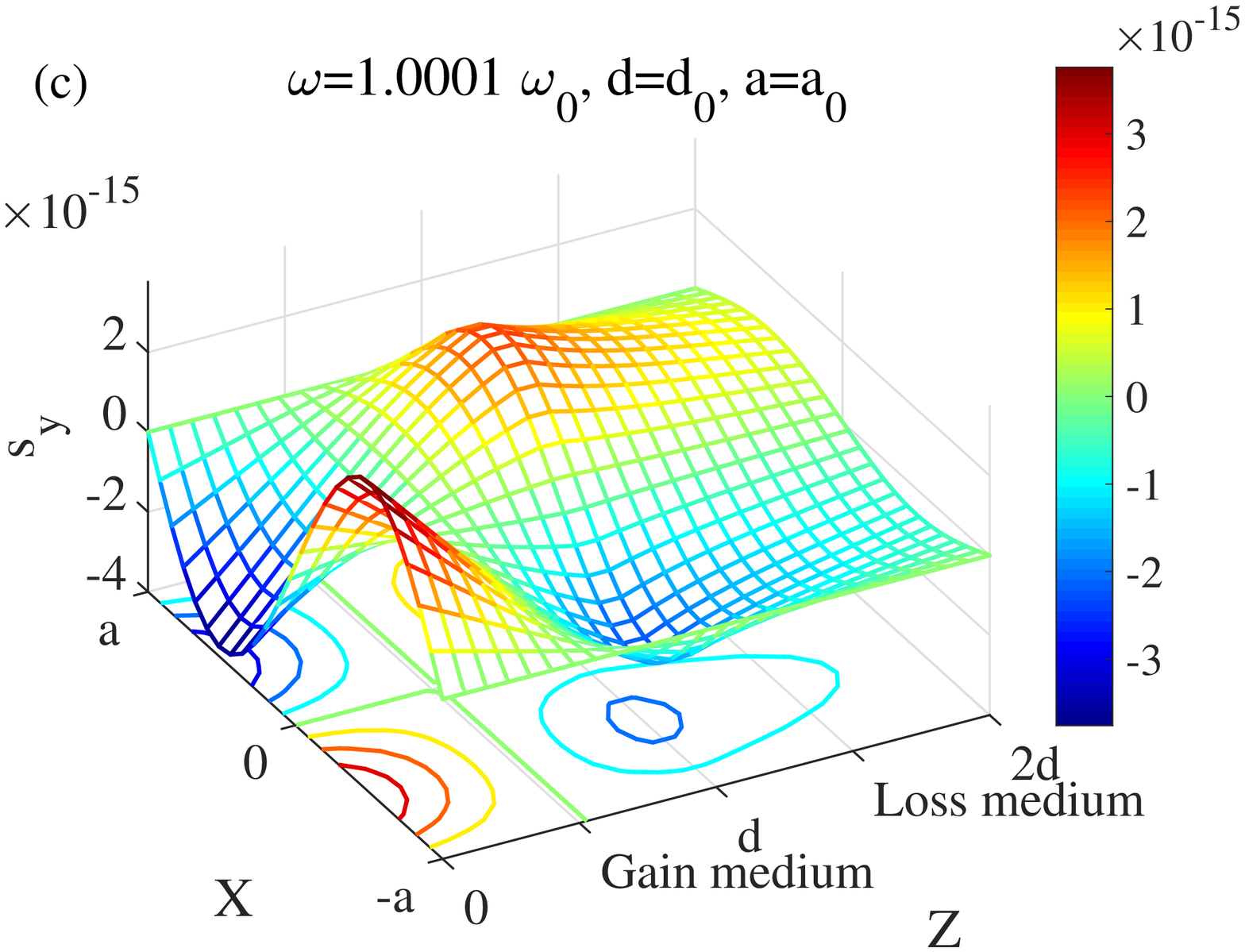}
	\caption{Surface plot of $S_y$ throughout the waveguide for (a) $\omega=0.9999\omega_0$, $d=d_0$, $a=a_0$ (b) $\omega=\omega_0$, $d=d_0$, $a=a_0$ and (c) $\omega=1.0001\omega_0$, $d=d_0$, $a=a_0$, Here $\omega/\omega_0=1$, $a_0=0.062045958076$ $\mu$m  and $d_0=1.004358293913$ $\mu$m correspond to the resonance condition}
	\label{fig4}
\end{figure}
\par
Next we calculate the transverse spin density using Eq.\ref{eq15}. The results are shown in Figs. \ref{fig4}(a), (b) and (c) for both off-resonant and on-resonant cases. In each of these figures we have shown the surface plots of $s_y$ along with the corresponding contour profiles over the whole guiding region. For the off-resonant case we have allowed $\omega/\omega_0$ to be away from unity, while the other parameters  like $a$ and $d$ ware taken to be same as corresponding to the resonant values. A comparison of these figures (vertical axis) clearly exposes the role of the spectral singularity in the enhancement of the transverse spin. Moreover, the distribution is mainly governed by the same for the fields due to the guided fundamental mode in the active guide. 

\begin{figure}[htp] 
	\centering
		\includegraphics[width=0.5\linewidth]{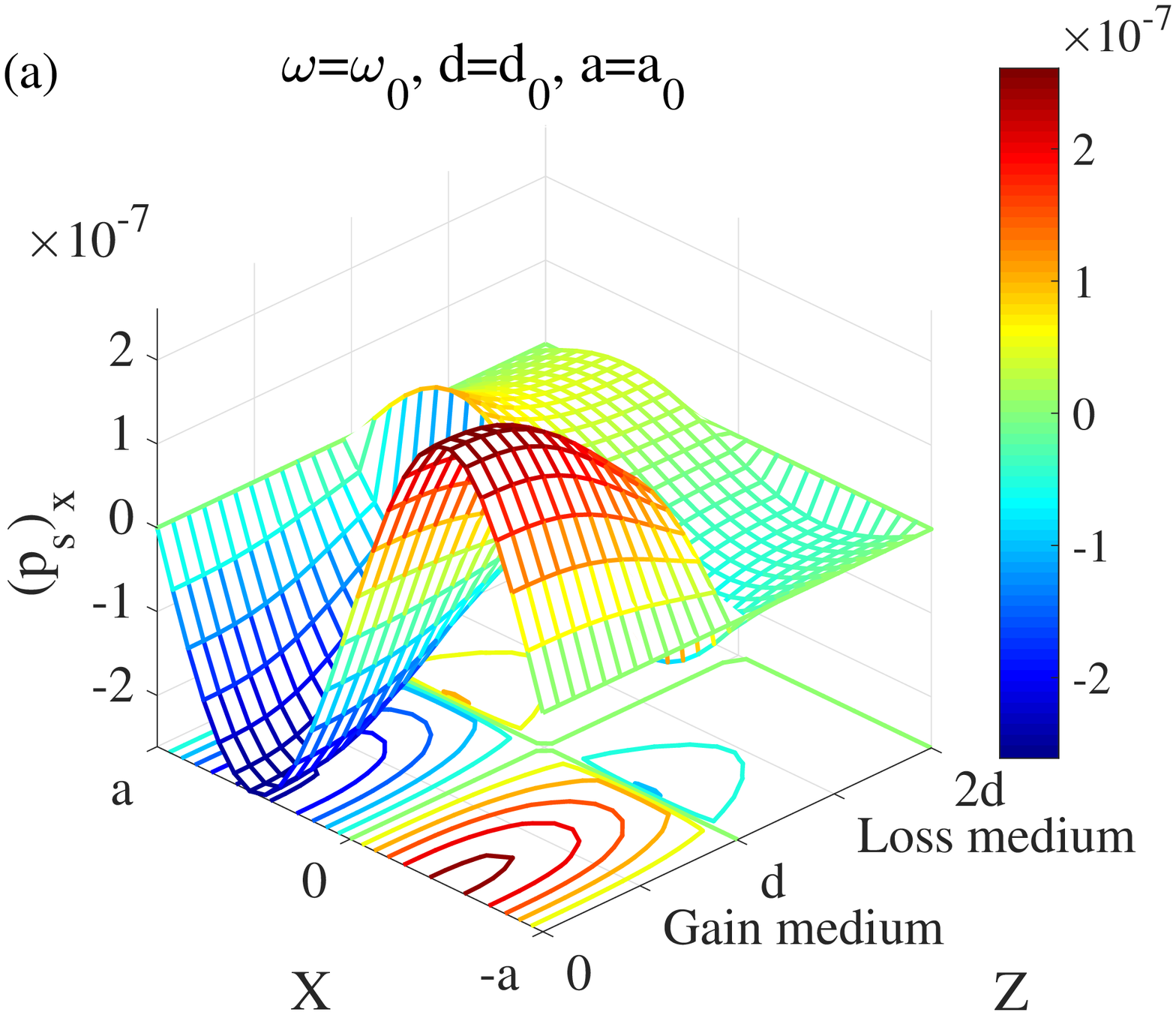}%
		\includegraphics[width=0.5\linewidth]{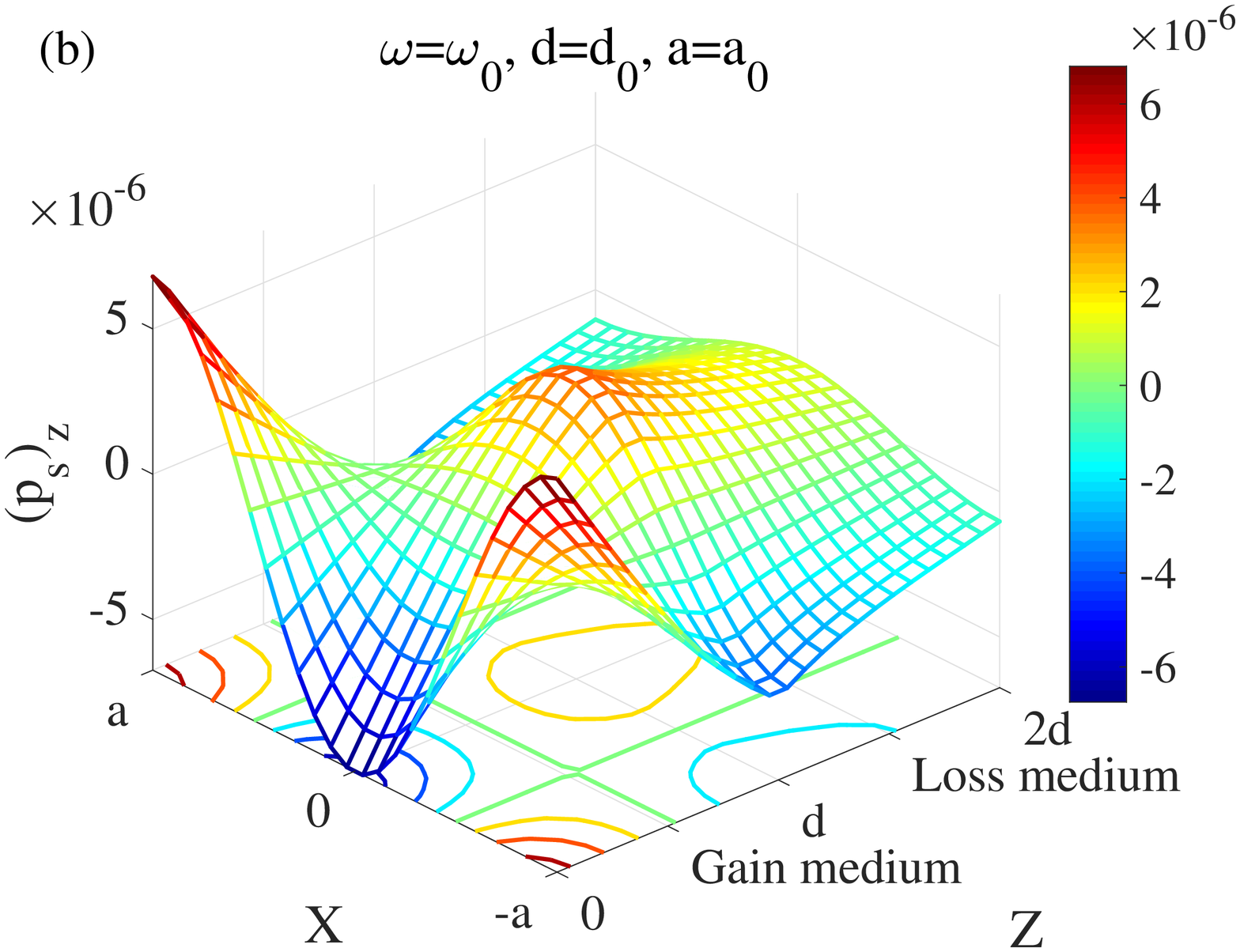}%
	\caption{Surface plot of (a) ${(p_s)}_x$ and (b) ${(p_s)}_z$ throughout the waveguide for $\omega = \omega_0$, $a = a_0= 0.062045958076$ $\mu$m and $d = d_0=1.004358293913$ $\mu$m}
		\label{fig5}
\end{figure}

\begin{figure}
	\begin{minipage}{.5\linewidth}
		\centering
		\includegraphics[width=\linewidth]{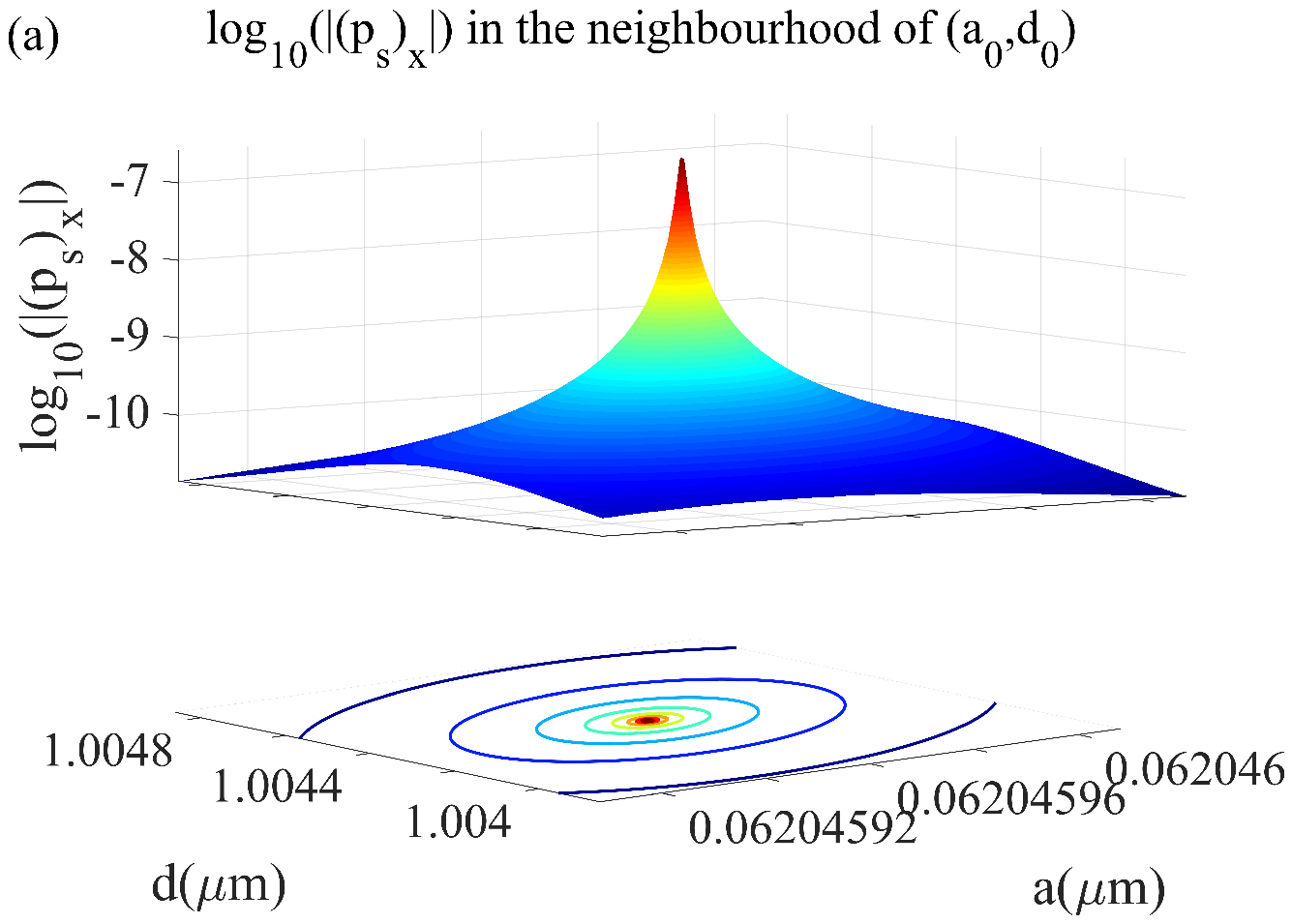}
	\end{minipage}%
	\begin{minipage}{.5\linewidth}
		\centering
		\includegraphics[width=\linewidth]{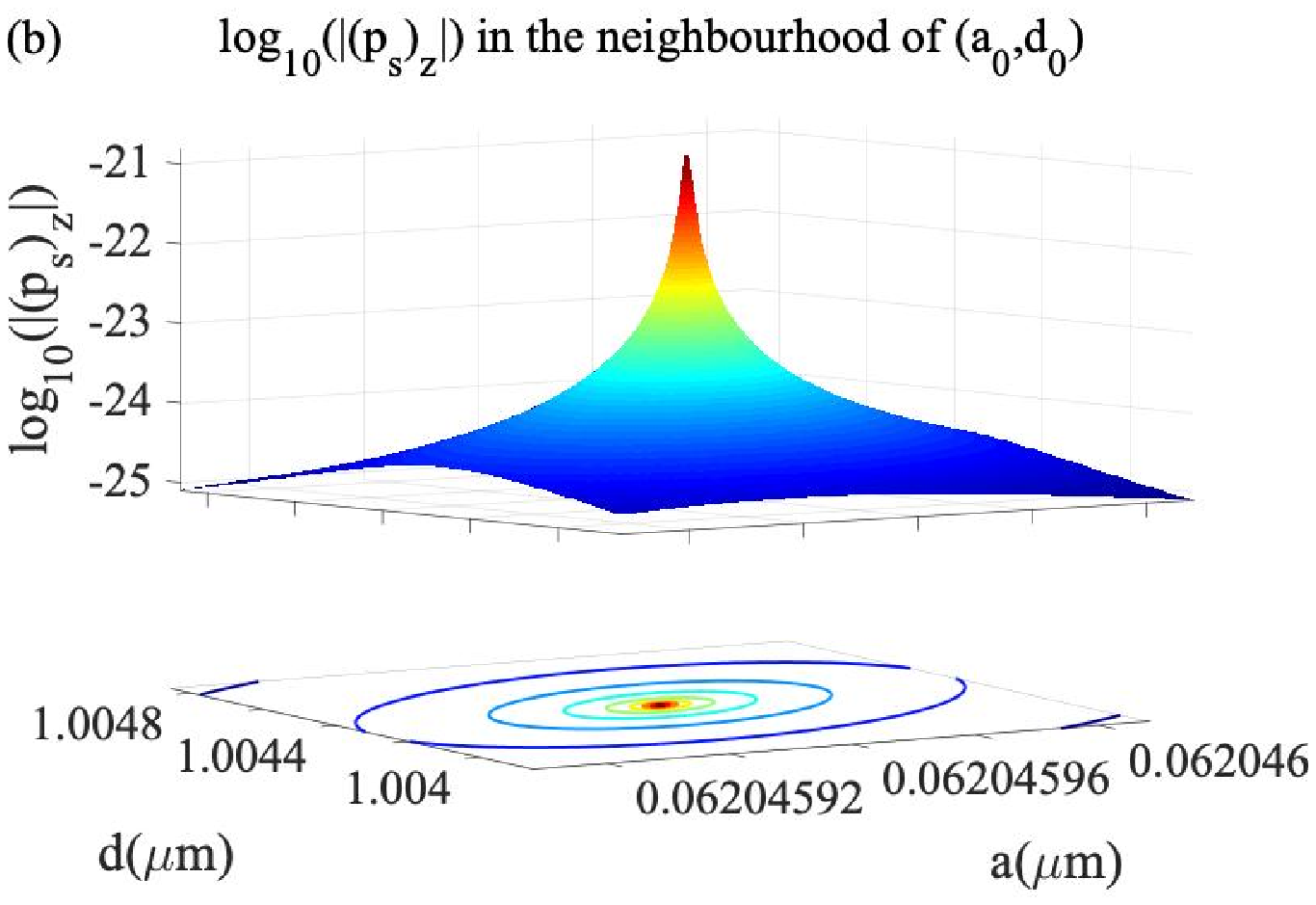}
	\end{minipage}\par\medskip
	\centering
	\includegraphics[width=0.5\linewidth]{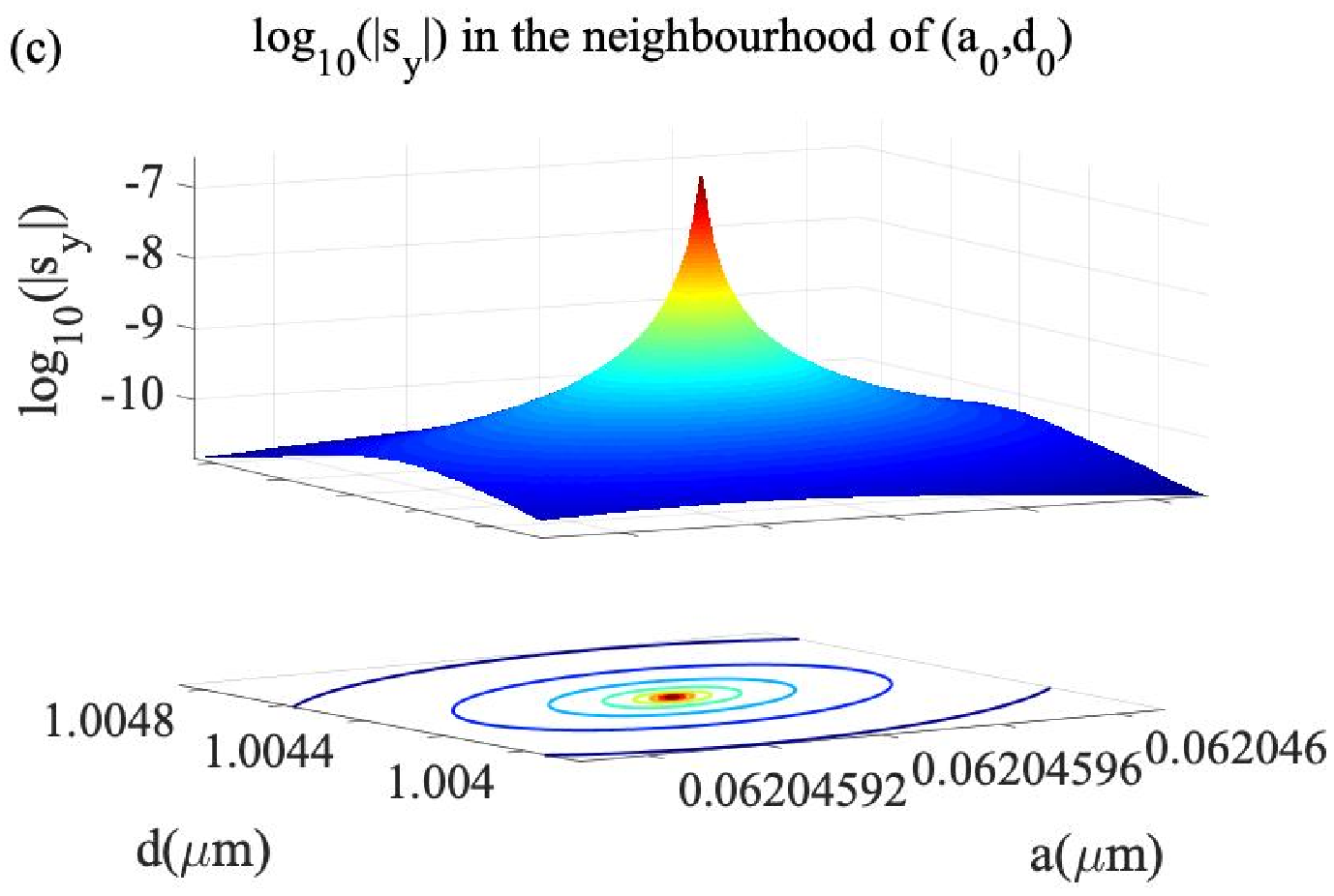}
	\caption{Variations of (a) $log(|(p_s)_x|)$, (b) $log(|(p_s)_z|)$ and (c) $log(|s_y|)$ as a function of a and d parameters of waveguide}
	\label{fig6}
\end{figure}

We next took at the dependence of the spin density on the system parameters. In standard scenarios of the origin of transverse spin in structured light there are contributions from a helicity dependent term as in the case with the evanescent waves. In fact this dependence is responsible for the spin locking phenomenon, which leads to the helicity dependent control of the transverse spin-momentum. In our case because of the linearly polarized character of the propagating modes, there is no helicity dependent contribution. However, because of the inhomogeneity of the medium (along $z$), there are both forward and backward propagating waves. The resulting $z$ dependence leads to the transverse $y-$ component of spin and the $x-$component of the Belinfante spin momentum. Note that in standard scenario with evanescent waves, in absence of helicity, the spin momentum density is purely longitudinal. In Fig.\ref{fig5}, we show the variation of $x-$ and $z-$ components of the spin momentum $\mathbf{(p_s)}$ across the waveguide at the spectral singularity.  It is clear that maximum/minimum values of the transverse and longitudinal components of the spin momentum is encountered at the input plane $z=0$. In Fig.\ref{fig6}, we show the logarithm of $|p_s|_x$, $|p_s|_z$ and $|s_y|$ at $\omega=\omega_0$ as a function of the waveguide parameters $a$ and $d$. It is observed that as $a\rightarrow a_0$ and $d\rightarrow d_0$, an enhancement of the transverse spin $|s_y|$ is achieved.
\par 
\section{Conclusion}
In conclusion, we have studied a $\mathcal{PT}$-symmetric gain-loss balanced waveguide and exploited its null-width resonance to enhance the transverse spin and the transverse spin momentum. It was shown that such a system has distinctive features as compared to the widely studied systems with evanescent waves. The transverse spin momentum was shown to have a component normal to the waveguide bounding planes. Our results clearly show the promises of active systems to enhance the elusive Belinfante spin momentum.

\bibliography{reference.bib}

\end{document}